\documentclass[aps,prb,twocolumn,amsmath,showpacs,amssymb]{revtex4}

\usepackage{graphics}
\usepackage{amsmath}
\usepackage{dcolumn}
\usepackage{citesort}

\begin{document}

\title{{\rm\small\hfill (submitted to J. Chem. Phys.)}\\
Representing molecule-surface interactions with symmetry-adapted neural networks}

\author{J\"org Behler}
\affiliation{Fritz-Haber-Institut der Max-Planck-Gesellschaft,
Faradayweg 4-6, D-14195 Berlin, Germany}

\author{S\"onke Lorenz}
\affiliation{Fritz-Haber-Institut der Max-Planck-Gesellschaft,
Faradayweg 4-6, D-14195 Berlin, Germany}

\author{Karsten Reuter}
\affiliation{Fritz-Haber-Institut der Max-Planck-Gesellschaft,
Faradayweg 4-6, D-14195 Berlin, Germany}

\date{\today}

\begin{abstract}
The accurate description of molecule-surface interactions requires a
detailed knowledge of the underlying potential-energy surface (PES).
Recently, neural networks (NNs) have been shown to be an efficient
technique to accurately interpolate the PES information provided for a
set of molecular configurations, e.g. by first-principles
calculations. Here, we further develop this approach by building the
NN on a new type of symmetry functions, which allows to take the
symmetry of the surface exactly into account. The accuracy
and efficiency of such symmetry-adapted NNs is illustrated by the
application to a six-dimensional PES describing the interaction of
oxygen molecules with the Al(111) surface.
\end{abstract}








\maketitle

\section{Introduction}

Molecule-surface interactions play a cental role in many
technologically relevant processes like heterogeneous catalysis,
semiconductor growth and corrosion. A detailed investigation of the
underlying elementary steps at an atomic scale, e.g. physisorption,
chemisorption, dissociation, diffusion, and desorption, is crucial
to gain the deeper understanding needed to identify new catalysts,
candidates for protective surface coatings, and chemically inert
surfaces. A central quantity in all these processes is the
potential-energy surface (PES), which gives the potential-energy of
the system as a function of the positions of the nuclei.
First-principles calculations, and in particular density-functional
theory (DFT), have become an important tool in providing accurate,
often quantitative information on PESs. Nevertheless, the
computational demands connected with the calculation of every single
PES point for systems involving an extended solid surface
still impose severe constraints, e.g. limiting the number and time span of
``on-the-fly'' {\em ab initio} molecular dynamics (MD) trajectories
to study the molecule-surface interaction. Consequently, important physical
quantities like sticking coefficients, which result from statistical averaging and which thus require e.g. a large number of MD runs to obtain converged
results, are still hardly accessible in this way.

As one possibility to determine such quantities from
first-principles, Gro\ss{} and Scheffler~\cite{gross95} introduced a
``divide and conquer'' approach, which allows to speed up the
calculation of the MD trajectories by splitting the problem into
three steps. First, the multi-dimensional PES is mapped on a finite
grid by calculating the energies for a number of configurations
using an accurate method like DFT. In a second step these points are
interpolated to a continuous PES representation by an
appropriate method that allows to access the energy and the forces
several orders of magnitude faster than the original DFT
calculations. In the mentioned example of MD simulations, the
forces at any nuclear configuration can then be obtained from the
interpolated PES representation at low computational cost,
thereby permitting the calculation of a sufficient number of MD trajectories.

Several approaches have been proposed for such interpolation
schemes. In analytical fits~\cite{gross95,gross98,wiesenekker96,wei98}
a reasonable functional form is ``guessed'' by physical intuition
and the free parameters of this functional form are optimized to
represent the set of DFT energies as
accurately as possible. If an appropriate functional form can be
found, this approach minimizes the required first-principles input
and avoids spurious, unphysical features in the PES. On the other
hand, with increasing dimensionality of the PES, the construction of
suitable functional forms becomes increasingly involved, and the
fixed functional form makes this approach also quite inflexible.
First attempts to optimize the functional form itself in the fitting
process by applying a genetic programming technique have hitherto
only been reported for up to three-dimensional
PESs~\cite{makarov98}. An alternative to analytic fits is the
modified Shepard interpolation~\cite{bettens99,crespos03,crespos04}.
In this method the potential close to a calculated DFT point is expanded as a
second-order Taylor series. The potential of a new configuration is
then constructed as a weighted sum over the Taylor expansions with
respect to the neighboring DFT points. The fitting procedure can be
further simplified by reducing the corrugation of the PES
~\cite{kresse00,busnengo00,crespos01}. For PESs describing a
molecule-surface interaction this can e.g. be achieved by
subtracting separately calculated PESs describing the interactions
of the individual atoms of the molecule with the surface. Finally,
we also mention an approach to efficiently represent PESs
by parameterizing a tight-binding Hamiltonian \cite{gross99}. In principle,
this approach requires only a relatively small number of DFT calculations
in order to obtain accurate fits to high-dimensional PESs. In practice, the
scheme is, however, unfortunately hampered by the indirect and cumbersome
tight-binding parameterization procedure.

In recent years, neural networks~\cite{hertz96} (NN) were also
successfully applied to fit PESs of small gas-phase
molecules~\cite{agrawal06,manzhos06,prudente98a,prudente98b,raff05,brown96,bittencourt03,no97}. Neural networks form a very general class of
functions~\cite{cybenko89,hornik89}, which in principle can
approximate any function to arbitrary accuracy without requiring any
information about the underlying functional form of the problem.
First applications to the description of molecule-surface
interactions have demonstrated the capabilities of this
approach~\cite{blank95,agrawal05,lorenz04,lorenz06}, but were
complicated by technical difficulties in achieving a proper
consideration of the symmetry of the solid surface~\cite{lorenz04}.
The aim of the present paper is therefore to overcome these limitations
by introducing a symmetry-adapted neural network representation of PESs
for molecule-surface interactions, which is based on a new type of
symmetry functions that takes the symmetry of the surface potential
exactly into account. The accuracy and efficiency of such a NN
representation, in particular for MD simulations, is demonstrated by studying
the interaction of oxygen with the Al(111) surface. For this, we concentrate
on the six-dimensional PES describing the interaction
with an O$_2$ molecule constrained to its spin-triplet state, which
could recently be shown to be of crucial importance in understanding
the low initial sticking coefficient reported
experimentally\cite{behler05,behler07a}.

\section{Neural Networks}

Inspired by the neural signal processing in biological
systems~\cite{mcculloch43}, the first artificial neural networks
have been introduced in 1959~\cite{rosenblatt58}. Since then NN
techniques have become a standard tool in many fields of research.
While they have been applied mainly to pattern recognition and
classification problems, neural networks also represent a very
general fitting scheme that in principle allows to approximate any
function to arbitrary accuracy~\cite{cybenko89,hornik89}. No
previous knowledge about the underlying functional form is required.
Instead, a number of known values of the function to be fitted is
presented to the NN in order to adapt a rather large number of
parameters using an optimization algorithm. Out of the many types of
neural networks, the class of multilayer feed-forward neural
networks~\cite{hertz96} has particularly proven to be a useful tool
for the representation of potential-energy
surfaces~\cite{blank95,lorenz04}.

\begin{figure}[t!]
\scalebox{0.6}{\includegraphics{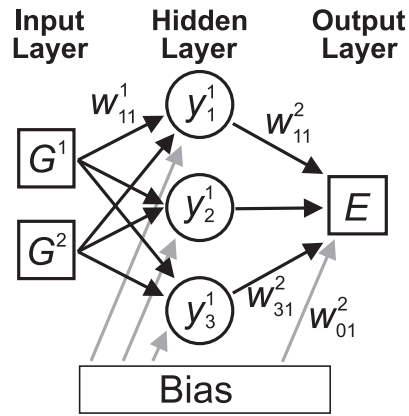}} \caption{Schematic structure
of a feed-forward neural network (NN). Two coordinates $\{G^{\mu}\}$
defining the molecular configuration are provided to the NN in the
nodes of the input layer. The total energy $E$ of the structure
calculated by the NN is given in the node of the output layer.
Between the input and the output layer is a hidden layer with three
nodes. All nodes are connected to the nodes in the adjacent layers
by weight parameters $w_{ij}^k$ that are optimized by fitting the
output energies to a provided training set of known input energies,
e.g. from density-functional theory calculations. The bias node acts
as an adjustable offset for nonlinear activation functions applied
at each node in the hidden and the output layer (cf. text).
\label{nnstructure}}
\end{figure}

The general structure of such a multi-layer feed-forward neural
network is shown schematically in Fig.~\ref{nnstructure}. It
consists of an input layer, one or more hidden layers and an output
layer. In each layer there is a certain number of nodes. When
representing a PES by a NN, the relevant coordinates of the system
determining the potential are fed into the input layer of the NN.
There are many possible choices for the input coordinates $G^{\mu}$
depending on the system to be studied. In the most simple case of
the one-dimensional interaction potential of an isolated diatomic molecule
the bond length would e.g. be an appropriate choice as input coordinate.
For molecules several bond lengths and angles can be used as input,
as has been shown for the PESs of several small
molecules~\cite{raff05,prudente98a}. The node in the output layer
provides the target quantity, i.e., in the present case the
potential energy of the system. However, NNs are not constrained to
fit only one quantity, and it would for example also be possible to
fit the potential and the forces acting on the atoms simultaneously.

In between the input and the output layer, there are one or more
hidden layers, each with a certain number of nodes. The term
``hidden layer'' indicates that the numerical values at the nodes of
these layers have no physical meaning and are just auxiliary
mathematical quantities. Each node in each layer is connected to the
nodes in the adjacent layers by so-called weights, the fitting
parameters of the NN. As illustrated in Fig. \ref{nnstructure}, the
weight parameter $w_{ij}^k$ is connecting node $j$ in layer $k$ with
node $i$ in layer $k-1$, where the input layer has the superscript
$k=0$. The value $y_j^k$ of node $j$ in layer $k$ is obtained from
the values $y_{i}^{k-1}$ of all nodes $i$ in the preceding layer
$k-1$ and from the connecting weights by
\begin{equation}
y_{j}^{k} \;=\; f_{\rm a}^k \left( w_{0j}^{k} + \sum_{i} w_{ij}^{k}y_{i}^{k-1} \right) \quad ,
\label{nodevalue}
\end{equation}
i.e., for each node the values of the nodes in the previous layer
are multiplied by the respective weights and added up to yield a
single number. On this number a so-called activation function
$f_{\rm a}^k$ is applied, which can be different for each layer.
Activation functions are typically sigmoidally shaped non-linear
functions, which introduce the capability to fit non-linear
functions into the NN. Frequently employed functional forms are the
hyperbolic tangent or Fermi-like functions. For very large or very
small arguments the activation functions converge to a constant
number, but for a certain interval the output changes significantly
in a non-linear way. In contrast, a linear activation function is
often employed in the output layer to avoid a constraint on the
possible range of output values. Finally, there is a bias weight
connected to each node in the hidden layers and the output layer,
which acts as an adjustable offset to shift the non-linearity regime
of the activation functions as needed to obtain an optimal fit.

The full analytic form for the small model NN shown in
Fig.~\ref{nnstructure} is given by the expression
\begin{equation}
E(\{G^{\mu }\}) = f_{a}^{2}\left(
w_{01}^{2}+\sum_{j=1}^{3}w_{j1}^{2}f_{a}^{1}\left(
w_{0j}^{1}+\sum_{\mu =1}^{2}w_{\mu j}^{1}G^{\mu }\right) \right)
\quad . \label{nnfunction}
\end{equation}
In general, each layer including input and output layers can contain
many more nodes than in the simple example shown in
Fig.~\ref{nnstructure}, and also more than one hidden layer is
typically used. The number of layers and nodes determines the
analytic form of the NN, and also analytic derivatives, i.e., the
forces, can be calculated. The overall NN architecture can then be
described following the scheme suggested in
Ref.~\onlinecite{lorenz06}. In this scheme, the NN in
Fig.~\ref{nnstructure} is a $\{2-3-1\quad tl\}$ network, where the
first number indicates the number of nodes in the input layer, the
last number the number of nodes in the output layer, and the
number(s) in between the number(s) of nodes in the hidden layer(s).
The employed activation function in each layer is labeled as $t$, if
a hyperbolic tangent is applied, or as $l$ for a linear function. In
the present example, a hyperbolic tangent is thus used in the hidden
layer, and a linear function in the output layer, hence $tl$.

In order to construct a continuous PES representation from a number
of known training energies (for example obtained from DFT), the
weight parameters of the NN are optimized in an iterative way to
reproduce these input energies as accurately as possible. Initially
the weights are chosen randomly, and for each known configuration
(i.e. point in the PES landscape) the potential predicted by the NN
is calculated. Taking these values and the corresponding original
input energies, an error function can be constructed. This error
function is then minimized to optimize the weight parameters and
thus requires the calculation of the derivatives of the output
energies with respect to each $w_{ij}^k$. Standard optimization
algorithms like steepest descent, which is called backpropagation in
the NN context, or conjugate gradient can be used to find the
optimum set of parameters. In particular for NNs also the extended
Kalman filter~\cite{shah92,blank94,bertsekas96} (EKF) has proven to
be a useful optimization tool. In the EKF the weights are adapted
not after the presentation of the full input data set, but after the
presentation of each single data point, which is intended to make the
optimization process less sensitive to local minima in the high-dimensional
optimization space.\cite{lorenz06}
Additionally, information from previous weight
updates is included by using a modified cost function including a
weighting factor for previous iterations. For details on the EKF we
refer to Ref.~\onlinecite{lorenz06} and references therein. Once an
optimal set of weight parameters has been found, the NN constitutes
the continuous PES representation, i.e. it can be employed to predict the
energy and forces at any point in the PES landscape.

{\em A priori} it is not known which network architecture will be
best for a given fitting problem, and tests are necessary to find
the optimum number of hidden layers and nodes, as well as the best
activation functions for a given system. Too few nodes in the hidden
layers will typically result in underfitting, i.e., important
features of the PES will be erroneously smoothed out. More nodes
increase the flexibility of the NN, but can lead to overfitting,
i.e., artificial features appear in the PES. In general, an
advisable strategy seems therefore to use the smallest possible NN
that yields the desired accuracy.

In order to check on overfitting, one can split the full input data
set into a training set used to optimize the NN parameters, and a
test data set, which is not employed in the fit. The set of NN
parameters that yields the lowest error for the test set may then be
interpreted as having the best predictive power for unknown
structures.

\section{Symmetry-adapted neural networks \label{symsection}}

The intention of the present work is to employ NNs to represent the
PES describing the interaction of a molecule approaching an extended
single-crystal surface. Addressing an important class of molecules,
we will focus our discussion on diatomic molecules, and employ
furthermore the so-called frozen-surface approximation. A possible
generalization of our approach is then discussed in Section \ref{discussion}.

In the frozen-surface approximation,
the substrate atoms are assumed to be fixed to their positions
pertinent to the clean surface. The dimensionality of the problem is
thereby reduced to the degrees of freedom representing the impinging
molecule, i.e. six dimensions for the case of diatomic molecules.
Instead of the Cartesian coordinates of the two atoms, usually the
center of mass coordinates $X$, $Y$, and $Z$ (where the $x$- and $y$-axes
are parallel to the surface, and the $z$ axis points out of plane), as
well as the molecular bond length $r$, the angle between the
molecular axis and the surface normal $\theta$, and the angle
between the projection of the molecular axis into the surface plane
and the positive $x$-axis $\phi$ are used to describe the molecular
configuration. These coordinates can be grouped into two classes:
the coordinates $X$, $Y$, $\theta$ and $\phi$, in which the
potential is periodic, and the ``non-periodic'' coordinates $Z$ and
$r$.

Due to the lateral symmetry of the solid surface it is sufficient to
map the PES only for molecular configurations inside the irreducible
wedge of the surface unit-cell. However, after the NN parameterization
it is usually desirable to also have the appropriate PES information
available outside this irreducible wedge. In MD simulations
intended to be run on the NN represented PES, the molecule should
for example not be constrained to a motion inside the irreducible
wedge only, but should be able to move from one wedge to another.
Consequently, the NN has to provide the energy and forces for all
possible molecular configurations also outside the
symmetry unique wedge. A straightforward solution would be to expand the
energetic input data to a wider range of lateral $X$ and $Y$ coordinates
by suitably copying the data inside the irreducible wedge into neighboring
wedges before the fit, and then to train the NN to a larger surface
area. Such a procedure is necessarily very inefficient due to the
increased input data set and larger NN size, i.e., number of
parameters that would be required. Furthermore, the NN fit is
necessarily only an approximation to the underlying energetic data,
and if parts of the PES, which are equivalent by symmetry, are fitted independently, the symmetry is numerically broken.

A solution to this problem has been suggested by Lorenz, Gro\ss{}
and Scheffler by introducing functions including the symmetry of the
surface.~\cite{lorenz04,lorenz06} Basically, the problem of
assigning energies to molecular configurations is separated into two
steps. In the first step the molecular coordinates are mapped on the
``symmetry functions'', which describe the symmetry of the surface,
and thereby the symmetry of the PES. The resulting symmetry
function values replace the original molecular coordinates as input
for the NN, and in a second step the energies are assigned to the
symmetry function values by the NN. This procedure ensures that the
symmetry properties of the symmetry functions are fully included in
the NN representation of the PES. The construction of these symmetry
functions is similar to the construction of the functional form in
an analytic fit~\cite{gross95}. Yet, the function values do not need
to yield the total energy of the system, but only its symmetry in
form of a set of function values.

Unfortunately, the hitherto proposed symmetry functions are valid only
for a given system and are difficult to construct even for
simple low-dimensional PESs, because they have to describe correctly
the complex interdependence of all ``periodic'' coordinates $X$, $Y$,
$\theta$ and $\phi$.~\cite{lorenz04,lorenz06} Up to now, such functions have correspondingly
been constructed using physical intuition, which can easily lead to
artificial or missing symmetries. Since this can have severe consequences
on the represented PES topology, and thereby on molecular trajectories
in MD simulations run on the NN interpolated PES, the present work
aims to improve on this situation by developing a more general scheme to
construct symmetry functions, which include the full symmetry of a
given system and are free of spurious symmetries not present in the
true molecule-surface interaction.

\subsection{Interaction of an atom with a fcc(111) surface}

\begin{figure}[t!]
\scalebox{1.1}{\includegraphics{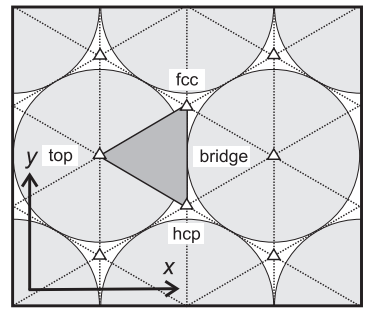}} \caption{Top view
illustrating the symmetry of a fcc(111) surface. The irreducible
wedge of the surface unit-cell (dark grey triangle) is spanned by
the top, fcc and hcp sites. The dotted lines represent mirror planes
and the white triangles show the positions of threefold rotation
axes perpendicular to the surface. The first layer substrate atoms
are shown as grey circles, and the coordinate system illustrates
the directions of the employed in-plane $x$- and $y$-axes.}
\label{Al111symmetry}
\end{figure}

\begin{figure}[t!]
\scalebox{0.3}{\includegraphics{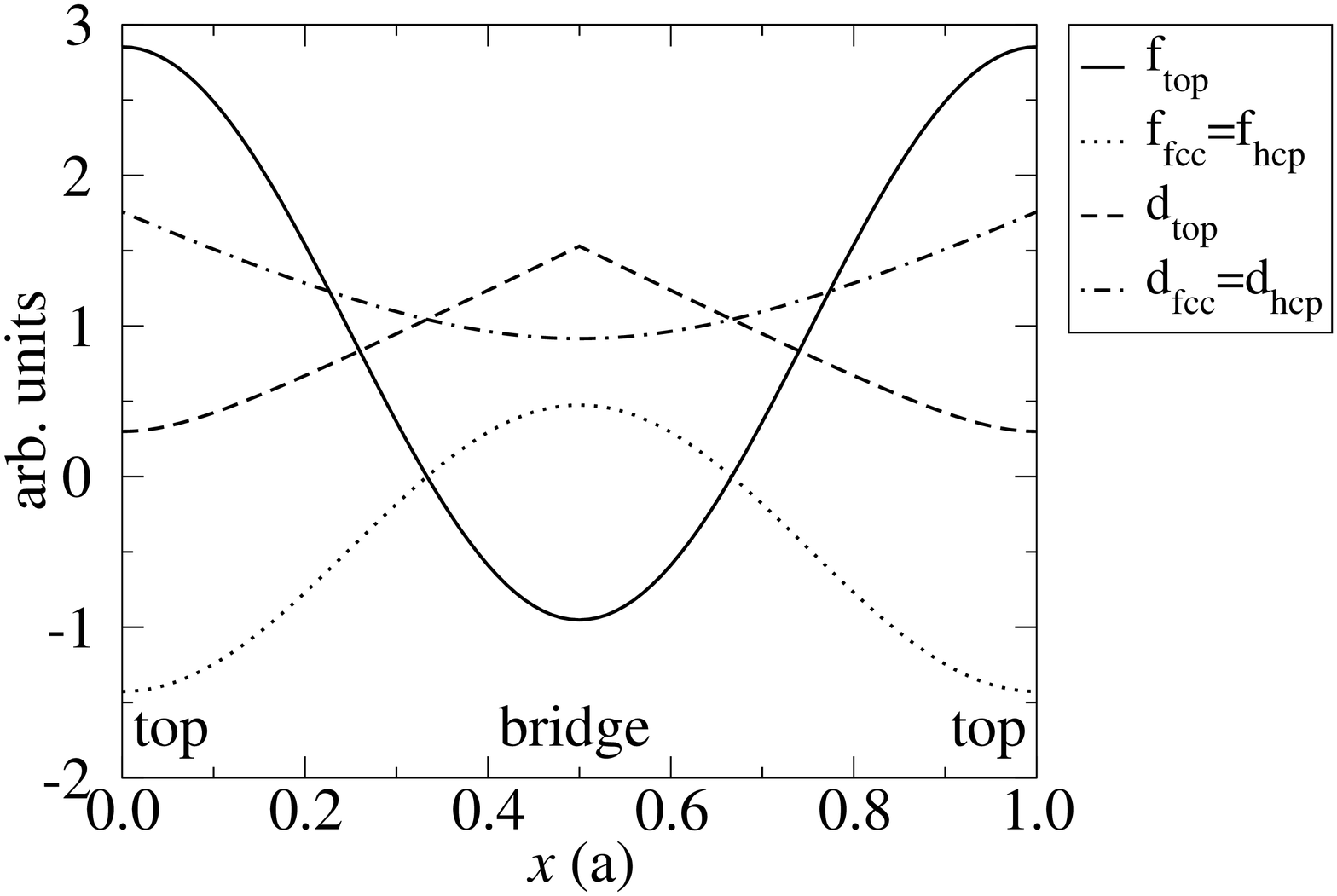}} \caption{Values of the atomic
distances $d_{\rm top}$, $d_{\rm fcc}$ and $d_{\rm hcp}$ of an atom
moving along the $x$-axis from the top site along the bridge site to
a neighbored top site, cf. Fig. \ref{Al111symmetry}. At the bridge
site $d_{\rm top}$ shows a discontinuity in the derivative, which is
not present in the atomic Fourier terms $f_{\rm top}$, $f_{\rm fcc}$
and $f_{\rm hcp}$ in Eqs.~(\ref{fouriertop}-\ref{fourierhcp}).}
\label{discontinuity}
\end{figure}

We begin developing our concept by looking
at the symmetry of the interaction of an atom with a fcc(111)
surface. A schematic top view explaining the irreducible wedge of
the surface unit-cell is shown in Fig. \ref{Al111symmetry}. The
position of the atom over the surface is uniquely defined by its
three Cartesian coordinates $X$, $Y$, and $Z$, where $X$ and $Y$ are
in the surface plane as indicated in Fig. \ref{Al111symmetry}. The
lateral symmetry of the surface could be considered in a
straightforward manner by replacing
these coordinates with the distances $d_{\rm top}$, $d_{\rm fcc}$
and $d_{\rm hcp}$ of the atom to the closest top, fcc and hcp
surface sites, i.e., the edges of the irreducible wedge of the
surface unit-cell. If in a MD simulation an atom crosses the border
of the symmetry unique wedge of the surface, the closest reference
surface sites change and the set of \{$d_{\rm top}$, $d_{\rm fcc}$,
$d_{\rm hcp}$\} values naturally incorporates the symmetry as shown in
Fig.~\ref{discontinuity} for an atomic motion from a top site via a
bridge site to a neighboring top site. Nevertheless, although describing the
periodicity of the surface correctly, the set \{$d_{\rm top}$,
$d_{\rm fcc}$, $d_{\rm hcp}$\} is not an appropriate choice for the
symmetry functions to fit the total energy of the system, since in
MD simulations also the derivatives of the energy with respect to
the atomic coordinates, i.e., the forces, are required. As can be
seen in Fig.~\ref{discontinuity} for the sample atomic motion,
$d_{\rm top}$ shows a discontinuity in its derivative at the wedge
boundary originating from the switch to another reference top site,
and consequently this discontinuity is also present in the forces.

\begin{figure*}[t!]
\scalebox{0.8}{\includegraphics{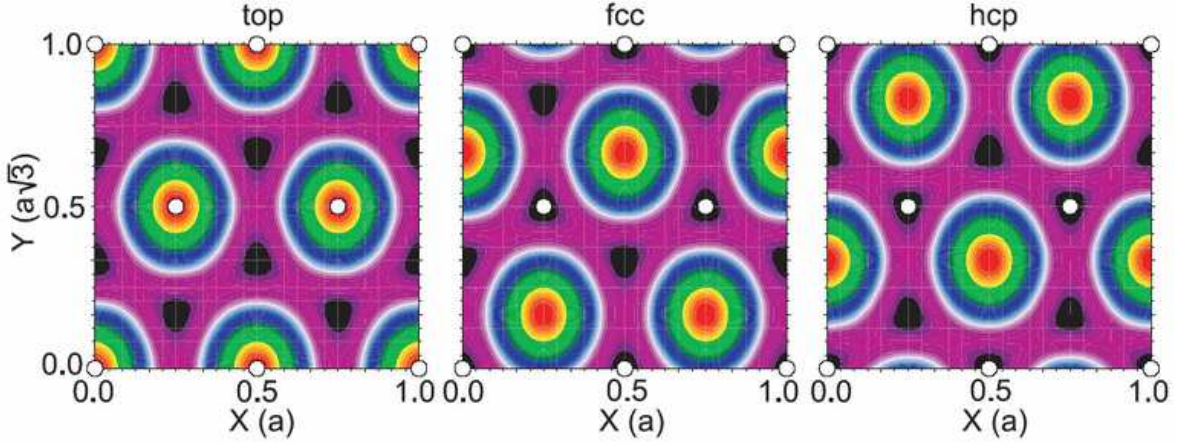}} \caption{(Color online)
Fourier terms for the interaction of an atom with a fcc(111)
surface. In (a) the top-term $f_{\rm top}$, in (b) the fcc-term
$f_{\rm fcc}$, and in (c) the hcp-term $f_{\rm hcp}$, defined in
Eqs. (\ref{fouriertop} - \ref{fourierhcp}) respectively, are shown as a function
of $x$ and $y$ in units of the lattice constant $a$. In all three
figures the white circles represent the position of the first layer
substrate atoms.} \label{fourieratom}
\end{figure*}

We solve this problem of discontinuities in the derivatives by
replacing the atomic distances by Fourier terms centered at the
high-symmetry sites, which describe the position of the atom in a
unique way as well. For a fcc(111) surface, these atomic Fourier terms are
\begin{eqnarray}
f_{\rm top} &=& \bigg[ \cos\left(\frac{2\pi}{a} \left(\left(X-X_{\rm
t}\right) +\frac{1}{\sqrt{3}}\left(Y-Y_{\rm
t}\right)\right)\right) \nonumber \\ &&+ \cos\left(\frac{4\pi}{a\sqrt{3}}\left(Y-Y_{\rm t}\right)\right) \nonumber \\
 & & + \cos\left( \frac{2\pi}{a}\left(\left(X-X_{\rm t}\right)-\frac{1}{\sqrt{3}}\left(Y-Y_{\rm t}\right)\right)\right)\bigg]
 \nonumber \\&&\cdot \exp{\left(-\frac{1}{2}Z\right)} \label{fouriertop}
\end{eqnarray}
\begin{eqnarray}
f_{\rm fcc} &=& \bigg[ \cos\left(\frac{2\pi}{a} \left(\left(X-X_{\rm
f}\right)+\frac{1}{\sqrt{3}}\left(Y-Y_{\rm f}\right)\right)\right)
\nonumber \\&&+\cos\left(\frac{4\pi}{a\sqrt{3}}\left(Y-Y_{\rm f}\right)\right) \nonumber \\
 & & + \cos\left( \frac{2\pi}{a}\left(\left(X-X_{\rm f}\right)-\frac{1}{\sqrt{3}}\left(Y-Y_{\rm f}\right)\right)\right)\bigg]
\nonumber\\ &&\cdot \exp{\left(-\frac{1}{2}Z\right)}
\label{fourierfcc}
\end{eqnarray}
\begin{eqnarray}
f_{\rm hcp} &=& \bigg[ \cos\left(\frac{2\pi}{a} \left(\left(X-X_{\rm
h}\right)+\frac{1}{\sqrt{3}}\left(Y-Y_{\rm
h}\right)\right)\right)\nonumber\\&& +
\cos\left(\frac{4\pi}{a\sqrt{3}}\left(Y-Y_{\rm h}\right)\right) \nonumber \\
 & & + \cos\left(\frac{2\pi}{a}\left(\left(X-X_{\rm h}\right)-\frac{1}{\sqrt{3}}\left(Y-Y_{\rm h}\right)\right)\right)\bigg]
 \nonumber\\&&\cdot\exp{\left(-\frac{1}{2}Z\right)} \quad ,
\label{fourierhcp}
\end{eqnarray}
where the positions of the top, fcc and hcp sites are given by
($X_{\rm t},Y_{\rm t}$), ($X_{\rm f},Y_{\rm f}$) and ($X_{\rm
h},Y_{\rm h}$), respectively, and $a$ is the lattice constant of the
surface unit-cell. The exponential term
$\exp{\left(-\frac{1}{2}Z\right)}$ takes the dependence on the
vertical distance $Z$ to the surface into account, and ensures in particular
that for a large atom-surface distance the function values and therefore
also the fitted energy should become independent of the actual
values of $X$ and $Y$. The Fourier terms are compared to the
simple distance terms in Fig.~\ref{discontinuity}. They have continuous
derivatives, which demonstrates that this function set is now a
suitable choice for fitting the PES of an atom interacting with a
fcc(111) surface, while naturally including the full symmetry of the
surface. This is also directly apparent from the full lateral periodicity
of the functions $f_{\rm top}$, $f_{\rm fcc}$ and $f_{\rm hcp}$ that
is plotted in Fig.~\ref{fourieratom}.

\subsection{Interaction of a diatomic molecule with a fcc(111) surface}

It is intuitive to generalize the concept of Fourier term based symmetry functions to molecules by building on the Fourier terms for each constituent atom of the molecule. For a diatomic molecule, the positions of both atoms
can then e.g. be described by Fourier terms like
\begin{eqnarray}
f_{\rm top1} &=& \bigg[C_1 - \cos\left(\frac{2\pi}{a}
\left(\left(X_1-X_{\rm t}\right)+\frac{1}{\sqrt{3}}\left(Y_1-Y_{\rm
t}\right)\right)\right)\nonumber \\ && - \cos\left(\frac{4\pi}{a\sqrt{3}}\left(Y_1-Y_{\rm t}\right)\right) \nonumber \\
 & & - \cos\left( \frac{2\pi}{a}\left(\left(X_1-X_{\rm t}\right)-\frac{1}{\sqrt{3}}\left(Y_1-Y_{\rm t}\right)\right)\right) +Z_1^2\bigg] \nonumber\\
 &&\cdot e^{-\frac{1}{2}Z} \label{fouriertop1}
\end{eqnarray}
\begin{eqnarray}
f_{\rm fcc1} &=& \bigg[C_1 - \cos\left(\frac{2\pi}{a}
\left(\left(X_1-X_{\rm f}\right)+\frac{1}{\sqrt{3}}\left(Y_1-Y_{\rm
f}\right)\right)\right)
\nonumber \\ &&-\cos\left(\frac{4\pi}{a\sqrt{3}}\left(Y_1-Y_{\rm f}\right)\right) \nonumber \\
 & & - \cos\left( \frac{2\pi}{a}\left(\left(X_1-X_{\rm f}\right)-\frac{1}{\sqrt{3}}\left(Y_1-Y_{\rm f}\right)\right)\right) +Z_1^2\bigg] \nonumber\\
  &&\cdot e^{-\frac{1}{2}Z} \label{fourierfcc1}
\end{eqnarray}
\begin{eqnarray}
f_{\rm hcp1} &=& \bigg[C_1 - \cos\left(\frac{2\pi}{a}
\left(\left(X_1-X_{\rm h}\right)+\frac{1}{\sqrt{3}}\left(Y_1-Y_{\rm
h}\right)\right)\right) \nonumber\\&&-
\cos\left(\frac{4\pi}{a\sqrt{3}}\left(Y_1-Y_{\rm h}\right)\right) \nonumber \\
 & & - \cos\left(\frac{2\pi}{a}\left(\left(X_1-X_{\rm h}\right)-\frac{1}{\sqrt{3}}\left(Y_1-Y_{\rm h}\right)\right)\right) +Z_1^2\bigg] \nonumber\\
 &&\cdot e^{-\frac{1}{2}Z} \label{fourierhcp1}
\end{eqnarray}
\begin{eqnarray}
f_{\rm top2} &=& \bigg[C_1 - \cos\left(\frac{2\pi}{a}
\left(\left(X_2-X_{\rm t}\right)+\frac{1}{\sqrt{3}}\left(Y_2-Y_{\rm
t}\right)\right)\right)\nonumber\\&& - \cos\left(\frac{4\pi}{a\sqrt{3}}\left(Y_2-Y_{\rm t}\right)\right) \nonumber \\
 & & - \cos\left( \frac{2\pi}{a}\left(\left(X_2-X_{\rm t}\right)-\frac{1}{\sqrt{3}}\left(Y_2-Y_{\rm t}\right)\right)\right) +Z_2^2\bigg]\nonumber\\
 &&\cdot e^{-\frac{1}{2}Z} \label{fouriertop2}
\end{eqnarray}
\begin{eqnarray}
f_{\rm fcc2} &=& \bigg[C_1 - \cos\left(\frac{2\pi}{a}
\left(\left(X_2-X_{\rm f}\right)+\frac{1}{\sqrt{3}}\left(Y_2-Y_{\rm
f}\right)\right)\right) \nonumber\\&&-
\cos\left(\frac{4\pi}{a\sqrt{3}}\left(Y_2-Y_{\rm f}\right)\right) \nonumber \\
 & & - \cos\left( \frac{2\pi}{a}\left(\left(X_2-X_{\rm f}\right)-\frac{1}{\sqrt{3}}\left(Y_2-Y_{\rm f}\right)\right)\right) +Z_2^2\bigg]\nonumber\\
 &&\cdot e^{-\frac{1}{2}Z} \label{fourierfcc2}
\end{eqnarray}
\begin{eqnarray}
f_{\rm hcp2} &=& \bigg[C_1 - \cos\left(\frac{2\pi}{a}
\left(\left(X_2-X_{\rm h}\right)+\frac{1}{\sqrt{3}}\left(Y_2-Y_{\rm
h}\right)\right)\right) \nonumber\\&&-
\cos\left(\frac{4\pi}{a\sqrt{3}}\left(Y_2-Y_{\rm h}\right)\right) \nonumber \\
 & & - \cos\left(\frac{2\pi}{a}\left(\left(X_2-X_{\rm h}\right)-\frac{1}{\sqrt{3}}\left(Y_2-Y_{\rm h}\right)\right)\right) +Z_2^2\bigg]\nonumber\\
&&\cdot e^{-\frac{1}{2}Z} \label{fourierhcp2}
\end{eqnarray}
which are functions of the Cartesian coordinates of the two atoms
$(X_1,Y_1,Z_1)$ and $(X_2,Y_2,Z_2)$. Apart from the constant $C_1$
to which we return below, the only modification compared to the atomic terms in
Eqs.~(\ref{fouriertop}-\ref{fourierhcp}), is the explicit addition
of the terms $Z_1^2$ and $Z_2^2$ to the respective Fourier terms
of atoms 1 and 2. This is necessary, because the multiplication
by the exponential term $\exp{\left(-\frac{1}{2}Z\right)}$ depending
only on the center-of-mass distance of the molecule from the surface
is not sufficient to distinguish different separations $Z_1$ and $Z_2$
of both atoms from the surface. This is resolved by the explicit
heights $Z_1$ or $Z_2$, and Eqs. (\ref{fouriertop1} - \ref{fourierhcp2}) then describe the positions of both atoms with respect to the top,
fcc and hcp sites uniquely.

What is still not defined, however, is the relative position of both atoms
to each other, since we obtain the same symmetry function values no matter
in which wedge at the surface, i.e., at which distance from each
other, the two atoms are located. This can be remedied by adding
a further symmetry function to the set, which is simply given by the
distance $r$ between both atoms of the molecule. The final set of symmetry
functions for a heteronuclear diatomic molecule is thus
\begin{subequations}\label{heterodiatom}
\begin{eqnarray}
G^{\prime,1} &=& f_{\rm top1} \\
G^{\prime,2} &=& f_{\rm fcc1} \\
G^{\prime,3} &=& f_{\rm hcp1} \\
G^{\prime,4} &=& f_{\rm top2} \\
G^{\prime,5} &=& f_{\rm fcc2} \\
G^{\prime,6} &=& f_{\rm hcp2} \\
G^{\prime,7} &=& r \quad .
\end{eqnarray}
\end{subequations}

\subsection{Incorporation of internal molecular symmetries}

The symmetry functions defined up to now allow to fully take the symmetry of
the solid surface into account, but do not exploit possibly existing
symmetries of the molecule itself. For diatomic molecules this would
be the case for homonuclear molecules like O$_2$, where an interchange
of both atoms should not change the symmetry function values and
therewith the input and output of the NN. Considering this
additional symmetry explicitly in the NN parameterization is obviously
desirable. Similar to the consideration of the surface symmetries, it
may even be mandatory from a numerical point of view to avoid artificial symmetry breaking and noise in the fitted total energies and forces.

Within the concept of Fourier term based symmetry functions, such
internal molecular symmetries may be accounted for by suitably
combining the atomic Fourier terms discussed up to now. For this,
it may be convenient to avoid negative function values and this
is the reason, why the arbitrary constant $C_1$ has been added to
the symmetry functions in Eqs. (\ref{fouriertop1} - \ref{fourierhcp2}).
In general, such a constant can be added at will, without changing
anything in the symmetry properties of the Fourier term. For a
heteronuclear diatomic molecule adding or not adding this constant
makes no difference, and one would simply choose $C_1 = 0$. However,
if one wants to avoid negative function values, a sufficiently positive
value for this constant may equally well be chosen.

\begin{figure*}[t!]
\scalebox{0.8}{\includegraphics{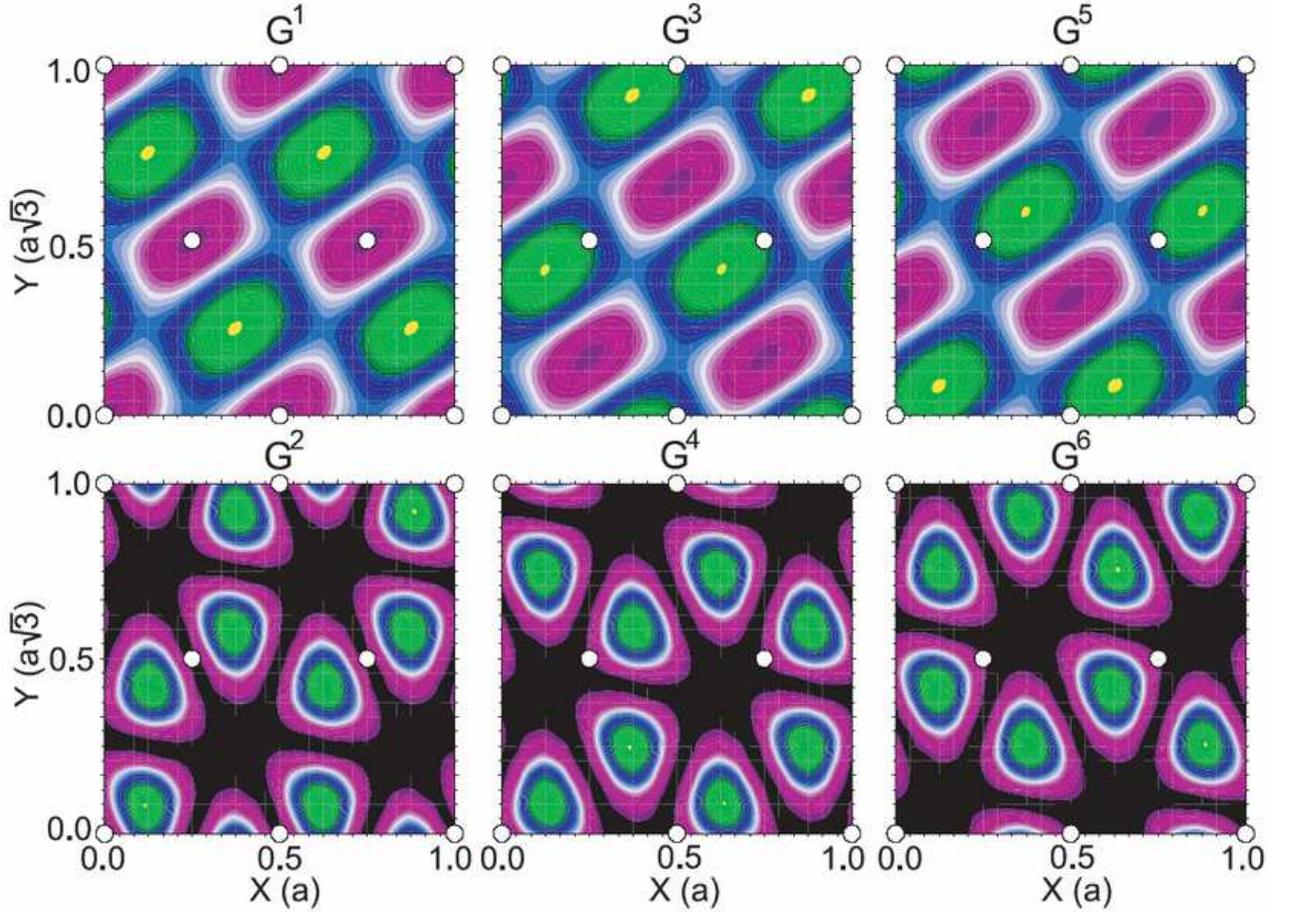}} \caption{(Color online) Plot
of the six symmetrized and anti-symmetrized Fourier terms of
Eq.~(\ref{symantisym}) as a function of the $X$ and $Y$ center of
mass coordinates of the molecule in units of the in-plane lattice
constant $a$. The positions of the top layer surface atoms are
marked by the white circles. In all plots the molecule has a
distance of 2.1 \AA{} from the surface, a bond length of 1.3 \AA{}
and an angular orientation of $\theta$ = 90$^\circ$ and $\phi$ =
30$^\circ$. The absolute function values have no meaning,
only the correct symmetry is required.} \label{moleculeplot}
\end{figure*}

For the example of a homonuclear diatomic molecule, the additional
symmetry with respect to exchange of the two atoms can then e.g. be
incorporated by symmetrizing and antisymmetrizing the Fourier terms
of both atoms for each surface site yielding the new set of symmetry
functions
\begin{subequations}\label{symantisym}
\begin{eqnarray}
G^{1} &=& \left( f_{\rm top1} + f_{\rm top2} \right)^2 \\
G^{2} &=& \left( f_{\rm top1} - f_{\rm top2} \right)^2 \\
G^{3} &=& \left( f_{\rm fcc1} + f_{\rm fcc2} \right)^2 \\
G^{4} &=& \left( f_{\rm fcc1} - f_{\rm fcc2} \right)^2 \\
G^{5} &=& \left( f_{\rm hcp1} + f_{\rm hcp2} \right)^2 \\
G^{6} &=& \left( f_{\rm hcp1} - f_{\rm hcp2} \right)^2 \\
G^{7} &=& r \quad .
\end{eqnarray}
\end{subequations}
These symmetrized and antisymmetrized atomic Fourier terms form
``molecular'' Fourier terms and still contain all structural
information. They are plotted in Fig.~\ref{moleculeplot} as a
function of the center-of-mass coordinates $X$ and $Y$ for fixed
values of $r$, $Z$, $\theta$ and $\phi$, exhibiting the correct
lateral periodicities of the solid surface.

\begin{figure*}[t!]
\scalebox{0.85}{\includegraphics{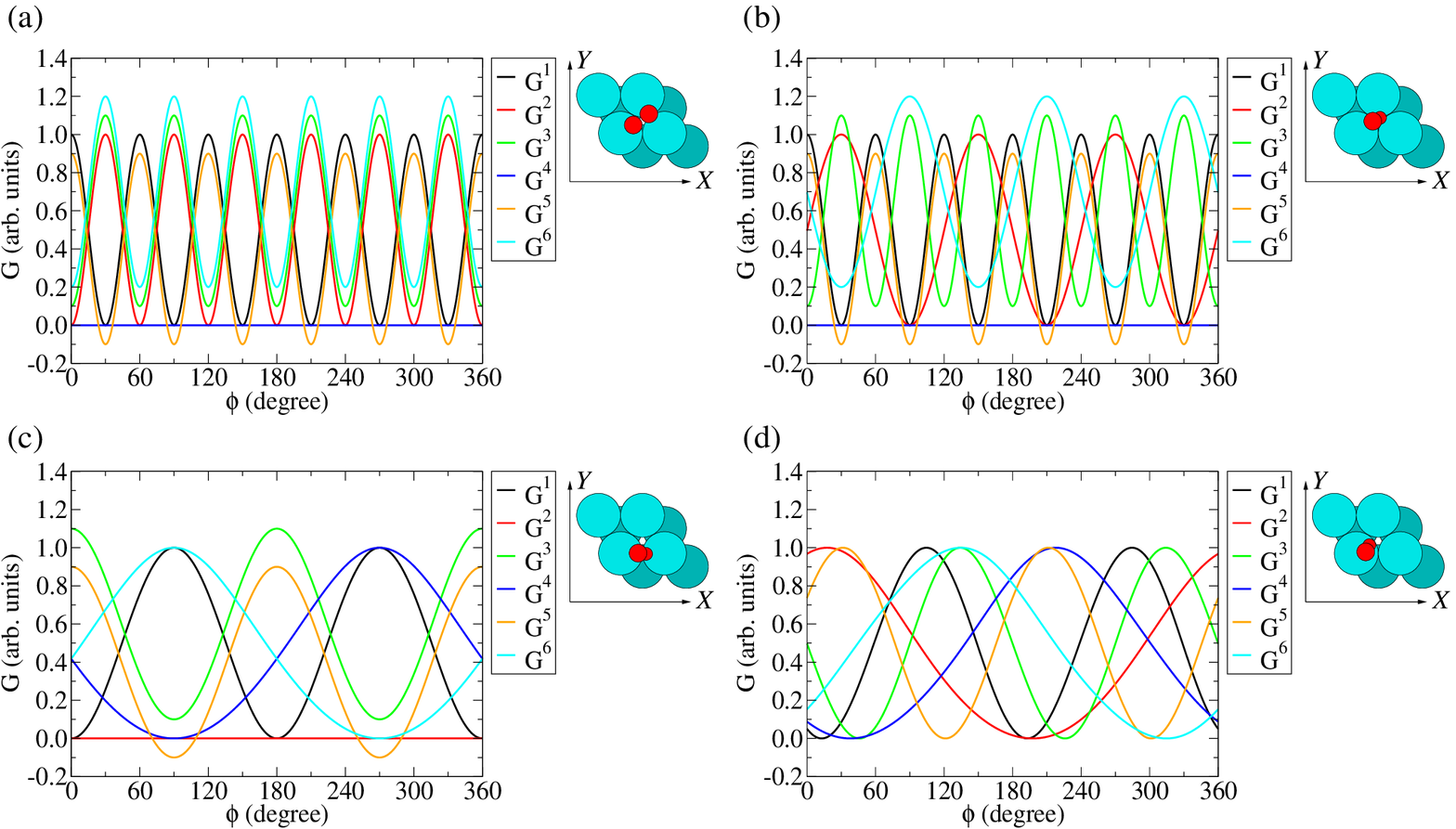}} \caption{(Color online)
Values of the symmetry functions $G^1 - G^6$ as defined in
Eq.~(\ref{symantisym}) for a molecular rotation about $\phi$ from
0$^\circ$ to 360$^\circ$, with $\phi$ being the angle between the
positive $x$ axis and the projection of the molecular axis into the
$xy$-plane. The insets show top views of the molecule above the
surface, illustrating the lateral position and angular orientation.
In (a) the molecule is parallel to the surface ($\theta$=90$^\circ$)
above a fcc site. In (b) the molecule is in a fcc site with an angle of $\theta$=30$^\circ$ to the surface normal.  In (c) the
molecule is above a bridge site with $\theta$=30$^\circ$. In (d) the molecule is
located at a low symmetry site ($X$=0.5a and
$Y$=$\frac{1}{24}\sqrt{3}$a) with $\theta$=30$^\circ$.
For plotting, the function values have been rescaled and shifted since the
absolute values have no meaning.}
\label{symmetryfuncplot}
\end{figure*}

The proper dependence on the angular orientation of the molecule is illustrated in Fig.~\ref{symmetryfuncplot}. Here, the symmetry function values are
plotted for a molecule rotating about $\phi$, i.e. for fixed coordinates $X$, $Y$, $Z$, $\theta$ and $r$. In Fig.~\ref{symmetryfuncplot}a the molecule is oriented parallel to the surface above an fcc site. A rotation by 60$^\circ$
transfers the molecule into an energetically equivalent
configuration. This is reflected in the set of symmetry function

values, which exhibit a periodicity of 60$^\circ$. If the molecule
is not parallel, but tilted with respect to the surface, the symmetry
is reduced from 6-fold to 3-fold. This is exemplified in
Fig.~\ref{symmetryfuncplot}b for $\theta$=30$^\circ$, and the
symmetry functions are able to describe this complex interdependence
between the two angles $\theta$ and $\phi$ correctly. Equivalent
performance of the symmetry functions is found for other
high-symmetry sites at the surface, namely the top and hcp sites.
Equally important, the reduced symmetry at other sites is also properly taken
into account, as shown in Fig.~\ref{symmetryfuncplot}c for a
molecule at a bridge site and tilted by $\theta$=30$^\circ$. The
symmetry of the remaining mirror plane at the bridge site is
reflected in the set of symmetry function values by the symmetry at
$\phi$=90$^\circ$ and $\phi$=270$^\circ$. Due to the inequivalence
of the hcp and fcc sites at the fcc(111) surface, there is no
symmetry with respect to $\phi$=0$^\circ$ and $\phi$=180$^\circ$ at
the bridge site. Finally, Fig.~\ref{symmetryfuncplot}d demonstrates
the capability of the symmetry functions for a low symmetry site at
the surface, namely for a molecule located at $Y=0.5$ and
$Y=\frac{1}{24}\sqrt{3}$. Here, the set of symmetry functions is
different for each value of $\phi$ because of the absence of
symmetry elements at this site.

\subsection{Adding redundant symmetry functions}

For diatomic molecules, the derived set of seven symmetry functions (either
$G^{\prime,1} - G^{\prime,7}$ for heteronuclear molecules or $G^{1} - G^{7}$
for homonuclear molecules) contains already exhaustive information on
the molecular configuration. However, adding further (redundant) symmetry
functions may nevertheless improve the numerical accuracy of the fit.
For this it is important to realize that the molecular coordinates
are always only mapped onto the symmetry functions, which in turn provide
the input for the NN. There is no need to ever invert this procedure,
i.e. to reconstruct the coordinates of each atom in the molecule that
correspond to a given set of symmetry function values. For the case
of a diatomic molecule, the six atomic coordinates may therefore be
mapped onto an arbitrarily large set of symmetry functions, if this
only helps to achieve a good NN fit to the PES. On the other hand,
one also has to recognize that the construction of such redundant symmetry
functions is necessarily system-specific, and the benefit of
including them in the set in terms of lowering the fitting error can
only be assessed by trial-and-error.

To provide an example for such redundant symmetry functions, we will
discuss below in the application to the dissociation of O$_2$ at Al(111)
that we found it useful to include a term depending only on
the molecule-surface separation
\begin{eqnarray}
G^8 &=& e^{-\frac{1}{2}Z} \quad ,
\end{eqnarray}
as well as three Fourier terms depending on the center of mass of
the molecule in the same way as the atomic terms in Eqs.~(\ref{fouriertop}-\ref{fourierhcp}).
\begin{eqnarray}\label{cmsfourier1}
G^9 &=& \bigg[ \cos\left(\frac{2\pi}{a} \left(\left(X-X_{\rm
t}\right)+\frac{1}{\sqrt{3}}\left(Y-Y_{\rm
t}\right)\right)\right)\nonumber\\&& + \cos\left(\frac{4\pi}{a\sqrt{3}}\left(Y-Y_{\rm t}\right)\right) \nonumber \\
 & & + \cos\left( \frac{2\pi}{a}\left(\left(X-X_{\rm t}\right)-\frac{1}{\sqrt{3}}\left(Y-Y_{\rm t}\right)\right)\right)\bigg]\nonumber\\&&\cdot\exp{\left(-\frac{1}{2}Z\right)}\\
\label{cmsfourier2}G^{10} &=& \bigg[ \cos\left(\frac{2\pi}{a}
\left(\left(X-X_{\rm f}\right)+\frac{1}{\sqrt{3}}\left(Y-Y_{\rm
f}\right)\right)\right)
\nonumber\\&& + \cos\left(\frac{4\pi}{a\sqrt{3}}\left(Y-Y_{\rm f}\right)\right) \nonumber \\
 & & + \cos\left( \frac{2\pi}{a}\left(\left(X-X_{\rm f}\right)-\frac{1}{\sqrt{3}}\left(Y-Y_{\rm f}\right)\right)\right)\bigg]\nonumber\\&&\cdot\exp{\left(-\frac{1}{2}Z\right)}\\
\label{cmsfourier3}G^{11} &=& \bigg[ \cos\left(\frac{2\pi}{a}
\left(\left(X-X_{\rm h}\right)+\frac{1}{\sqrt{3}}\left(Y-Y_{\rm
h}\right)\right)\right)
\nonumber\\&& + \cos\left(\frac{4\pi}{a\sqrt{3}}\left(Y-Y_{\rm h}\right)\right) \nonumber \\
 & & + \cos\left( \frac{2\pi}{a}\left(\left(X-X_{\rm h}\right)-\frac{1}{\sqrt{3}}\left(Y-Y_{\rm h}\right)\right)\right)\bigg]\nonumber\\&&\cdot\exp{\left(-\frac{1}{2}Z\right)}
\end{eqnarray}
These terms do not break the symmetry of the functions $G^1 - G^7$
(or $G^{\prime,1} - G^{\prime,7}$) and are motivated by the fact that
for a small almost spherical diatomic molecule the center of mass
position will have a pronounced influence on the energy expression.
Accordingly, although Eqs.~(\ref{cmsfourier1}-\ref{cmsfourier3})
contain redundant information, this does not complicate the fit,
but instead ``assists'' the NN in extracting the energetically relevant
structural information.

\subsection{Calculation of forces}

In order to perform molecular dynamics simulations the forces acting
on the atoms are required. The force $F_{\alpha}$ acting in the
direction of the molecular coordinate $\alpha$ can be obtained
analytically from the neural network by
\begin{eqnarray}
F_{\alpha}&=&-\frac{\partial E}{\partial \alpha}=-\sum_{\mu}\frac{\partial E}{\partial G^{\mu}}\cdot\frac{\partial G^{\mu}}{\partial \alpha} \quad .
\end{eqnarray}
The derivative of the total energy with respect to the symmetry
functions, i.e., the input of the NN, is determined by the network
structure. The derivative of the symmetry functions with respect to
the molecular coordinates is given by the definitions of the
symmetry functions, which thus need to have continuous derivatives.
Both can be evaluated analytically, and using these analytic derivatives it
is guaranteed that the forces are consistent with the energies, and
in particular, that the forces are zero at local minima of the
potential-energy surface. We found that the forces obtained from the
neural network are much more precise with respect to the symmetry of
the problem than forces directly obtained from DFT, since in the
latter case even in highly converged calculations the symmetry of
the forces is often broken by numerical noise.

\section{Application to O$_2$ dissociation at Al(111)\label{o2al111}}

We illustrate the use of the developed symmetry functions in the
application to the O$_2$ dissociation at an Al(111) surface. Here,
the six-dimensional PES representing the interaction of an O$_2$
molecule constrained to its spin-triplet state has been of
particular interest to explain the experimentally measured low
sticking coefficient. For further details on the physics of this
system, we refer to Refs.~\onlinecite{behler05} and
\onlinecite{behler07a}. Here this particular PES is simply taken as an
example to illustrate the NN interpolation scheme.

\subsection{Mapping of the six-dimensional PES}

As input data to the NN, the six-dimensional PES has been mapped
using density-functional theory as implemented in the DMol$^3$
code~\cite{delley90,delley00} and employing the RPBE
functional~\cite{hammer99} to describe electronic exchange and
correlation. The Kohn-Sham orbitals are expanded in a basis of
numerical atomic orbitals and polarization
functions~\cite{delley90}, and the spatial extent of the orbitals is
confined by a cutoff of 9 Bohr. A mesh of (3$\times$3$\times$1)
k-points has been used to sample the Brillouin zone. A Fermi
broadening of 0.1~eV has been applied to improve convergence and the
energy was subsequently extrapolated to 0~K. The spin-triplet on the
oxygen molecule is enforced by employing a locally-constrained DFT
approach described in detail in
Refs.~\onlinecite{behler05,behler07b}. A total of 3768 DFT data
points has been calculated, mostly contained in 38 different,
so-called ``elbow-plots'', which give the PES as a function of the
molecular bond length $r$ and distance to surface $Z$ for fixed
molecular orientation and surface site. The energy zero for the PES
has been defined for an infinitely large molecule-surface
separation, i.e., as the sum of the total energy of the clean
Al(111) surface and a free O$_2$ molecule at its equilibrium bond
length.

A detailed account of the DFT calculations and selection of the data
points can be found in Ref. \onlinecite{behlerdiss}. Here, we only
mention one aspect that was found to be of importance for the NN
fit. For small molecular bond-lengths and small molecule-surface
separations the potential becomes highly repulsive. This steep rise
in the energy (and consequently larger energy range to be fitted)
caused problems to the achieved accuracy of the NN fit. However,
these high-energy regions are not accessible in MD simulations
which typically focus on a maximum molecular kinetic energy of 1~eV, and consequently a
highly accurate mapping of these parts of the PES is not required. We thus applied a cutoff
function to constrain the highest potential energies to an energy
threshold of $E_{\rm t}=5$~eV using
\begin{eqnarray}
E=\Bigg\{ \begin{array}{rcl} E & \mbox{for} & E \leq E_{\rm t}-\Delta E \\
    E_{\rm t}-\exp\left({E_{\rm t}-\Delta E-E}\right) & \mbox{for}  & E> E_{\rm t}-\Delta E\end{array}
\end{eqnarray}
where $\Delta E$ has been set to 1~eV. The potential energies up to
+4~eV are thereby unmodified, while only the higher energies are
quenched to approach $E_{\rm t}$ asymptotically. Constraining the
highest energy of the PES in this way reduces the energy range to be
fitted, and was found to yield more accurate NN fits in the
remaining (relevant) energy range.

\subsection{NN optimization procedure}

\begin{table*}
\begin{tabular}{l|r}
\hline \hline
Characterization of data points& Weight\\ \hline
 $E < $1 eV  &  2.0\\
 $d$ = 1.224 \AA{} $\wedge $ (2.8 \AA{} $\leq$ $Z <$ 4.8 \AA )&1000.0\\
 $d$ = 1.224 \AA{} $\wedge $ $Z >$ 4.8 \AA&  5000.0\\
 (1.21 \AA{} $< d <$ 1.49 \AA) $\wedge$ (1.5 \AA{} $\leq Z <$ 2.6 \AA{})  $\wedge$ (-1.0 eV $< E <$ +1.0 eV)  & 11.0 \\
$d$ = 1.3 \AA{} $\wedge $ $Z$ = 2.1 \AA{} & 1067.0 \\ \hline \hline
\end{tabular}
\caption{\label{weights} Set of fitting weights assigned to the DFT
data points used in the NN optimization. The large weights enforce a
proper reproduction of the most relevant regions of the PES close to
the minimum energy paths, which are characterized by low energies or
bond lengths around the optimized free O$_2$ bond length for large
surface separations.}
\end{table*}

From the total of 3768 DFT energy points, 96 randomly chosen points
were used as independent test set not included in the NN
optimization procedure. Different ``fitting weights'' (not to be
confused with the weight parameters $w_{ij}^k$ connecting the NN
nodes) were assigned to the remaining DFT data points used to train
the NN. The motivation for this was the requirement that the PES
representation needs to be most accurate for the low-energy regions
along the minimum energy paths. DFT data points corresponding to
these regions were thus emphasized in the fit by higher fitting
weights, so as to enforce a most accurate reproduction by the
optimized NN. After several tests, the fitting weights compiled in
Table~\ref{weights} were chosen for the NN optimization.

In order to demonstrate the role of the symmetry functions, we
proceed in two steps. First, fits using only functions $G^1 - G^7$
are constructed, which already contain all relevant structural
information. In the next step fits using all 11 symmetry functions
$G^1 - G^{11}$ are constructed and the effect of the additional, redundant
functions on the numerical accuracy of the results is investigated.
Two aspects of the resulting PES representations are of interest:
First, the PES should be qualitatively correct, i.e., all symmetry
features of the system must be present. Second, also the
quantitative accuracy of the energies should be as close as possible
to the underlying DFT data.

\begin{figure*}[t!]
\scalebox{0.7}{\includegraphics{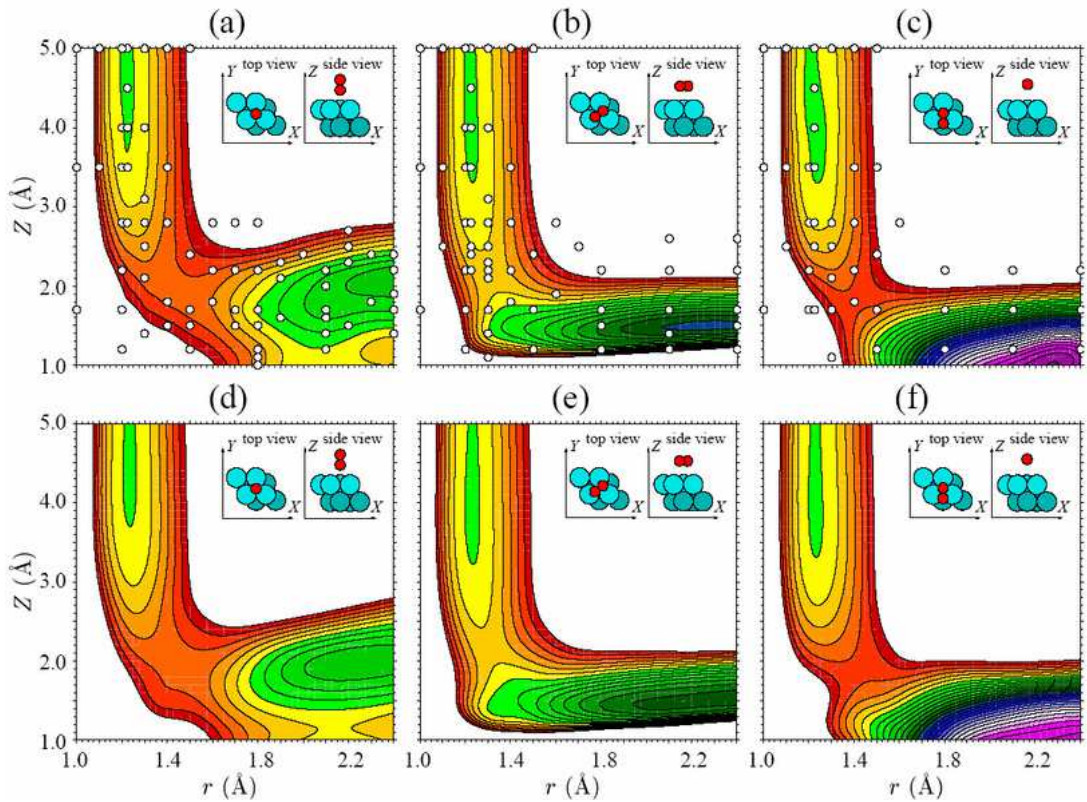}} \caption{(Color online)
Two-dimensional cuts (``elbow plots'') through the six-dimensional
spin-triplet potential-energy surface for the oxygen dissociation at
the Al(111) surface. The energy is shown as a function of the
center-of-mass distance of the molecule from the surface $Z$ and the
O$_2$ bond length $r$. In (a), (b) and (c) the elbow plots as
obtained with DFT (white data points) and splined within the two
dimensions are shown for the three different molecular orientations
described in the insets. In (d), (e) and (f) the corresponding elbow
plots obtained from the optimized NN potential are shown. Contour
lines indicate energy differences of 0.2 eV.} \label{elbowplots}
\end{figure*}

All in all 64 different NN architectures with varying numbers of
hidden layers, nodes per layer and parameters of the Kalman
filter~\cite{kalmandetails} have been tested. The best fit was
obtained using two hidden layers with 40 nodes per layer, the
hyperbolic tangent as activation function in the hidden layers and a
linear activation function in the output layer. Using symmetry
functions $G^1 - G^7$ only, i.e. the minimum structural information,
the root mean square errors (RMSE) of the training and the test sets
are 0.472~eV and 0.329~eV. The mean average deviations (MAD) are
0.166~eV and 0.196~eV. Due to the higher fitting weights assigned to
the low energy points (cf. Table~\ref{weights}) the accuracy of the
fit is much better for these DFT data points, which represent the
most important PES regions for the MD applications. The MAD for the
points with $E < 1.0$~eV is only 0.070~eV and the points along the
entrance channel having the O$_2$ equilibrium bond length of
$r=1.22$~\AA{} have a particularly small error of only 0.019~eV,
which is very important for an accurate description of steering
effects acting on slow molecules approaching the surface.

Employing all symmetry functions $G^1 - G^{11}$ the RMSE of the full training
and the test sets are reduced by one order of magnitude, to 0.049~eV
and 0.070~eV, respectively. The MADs are 0.023~eV and 0.033~eV. This
indicates that including the functions describing the center of mass
of the O$_2$ molecule strongly supports the fitting process for the
nearly spherical O$_2$ molecule. The fitting error for the important
low-energy points with $E < 1.0$~eV is now reduced to 0.012~eV and
the points along the entrance channel having the O$_2$ equilibrium
bond length of $r=1.22$~\AA{} have a very small error of only
1.4~meV. This excellent overall reproduction of the PES by the
optimized NN can also be seen by the three sample ``elbow-plots''
shown in Fig. \ref{elbowplots}. However, the inspection of
two-dimensional cuts through the 6-dimensional PES is not sufficient
to ensure an accurate representation of the PES in all dimensions.
We therefore confirmed the accuracy of the NN PES by comparing several MD
trajectories run on the NN PES with corresponding direct on-the-fly {\em ab initio} MD
trajectories. After 0.2 ps, which is about the time required by a
molecule with a translational kinetic energy of 0.15~eV starting at
$Z$=5~\AA{} to approach the barrier region at $Z$=2.5~\AA{}, the
positions in the NN PES MD and on-the-fly ab initio MD differed in all
cases by less than 0.1~\AA{}.

\begin{figure}[t!]
\scalebox{0.3}{\includegraphics{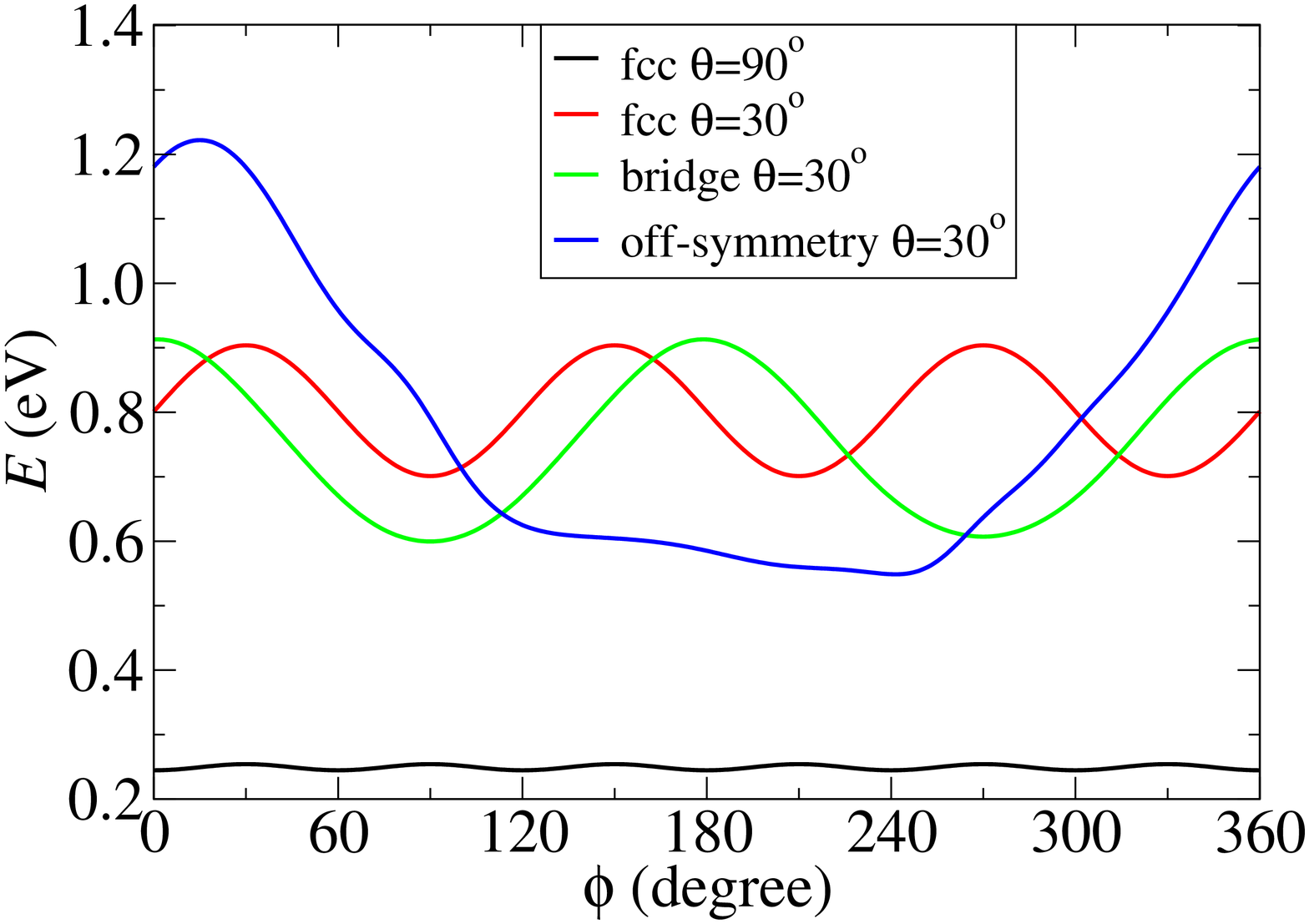}} \caption{(Color online)
Symmetry of the NN potential-energy for a molecular rotation about
$\phi$ at different surface sites corresponding to the plots of the
symmetry functions in Fig.~\ref{symmetryfuncplot}. The molecule
above the fcc site with $\theta=90^\circ$ shows a six-fold symmetry
like the set of symmetry function values in
Fig.~\ref{symmetryfuncplot}a. For $\theta=30^\circ$ a three-fold
symmetry is present like in Fig.~\ref{symmetryfuncplot}b, while
above the bridge site there is the symmetry of
Fig.~\ref{symmetryfuncplot}c. At the off-symmetry site there is no
symmetry like in Fig.~\ref{symmetryfuncplot}d.} \label{esymmetry}
\end{figure}

Apart from this quantitative assessment of the fitting accuracy, the
central aim of the present work is to provide a scheme to construct
NN potentials with the correct symmetry properties. In
Fig.~\ref{esymmetry} the potential obtained using functions $G^1 -
G^{11}$ is plotted for the molecular rotations described in
Fig.~\ref{symmetryfuncplot}. Clearly, the potential exhibits the
correct symmetry properties of the different surface sites.

\section{Discussion \label{discussion}}

The excellent representation achieved in the model case
O$_2$/Al(111) demonstrates that the proposed symmetry-adapted NNs
are very well capable of accurately interpolating a given set of
input PES data points. We acknowledge that mapping the molecular
coordinates on slightly involved symmetry functions, as well as
evaluating the NN output and its derivative on-the-fly in a MD
simulation is computationally slightly more demanding than employing
simple analytical potentials. Nevertheless, evaluation of the NN PES
is still about 5-6 orders of magnitude faster than direct DFT-based
MD simulations, and is less susceptible to PES representation errors
than the less flexible analytical potentials.

A clear disadvantage of the present NN interpolation scheme is that
consideration of further degrees of freedom requires a completely
new NN optimization. Whereas it is for example straightforward to consider
substrate motion in on-the-fly {\em ab initio} MD simulations, this
requires a mapping and interpolation of a correspondingly higher
dimensional PES in the NN scheme. One of us is presently developing
a new NN approach to this problem \cite{behler07c}, but the limited
dimensionality that can be well treated and the ensuing limitation to the
frozen-surface approximation are certainly the biggest drawbacks of
the NN scheme as presented here. In this respect, it is only a weak
justification, that consideration of only the molecular degrees of
freedom has proven to be a valid approximation for a number of
adsorbate systems.

With this in mind, the systematic construction of symmetry functions
based on atomic Fourier terms does improve on present NN schemes and
replaces the cumbersome empirical construction of complex symmetry
functions depending simultaneously on many degrees of freedom. The
correct choice of symmetry functions is essential for an accurate
representation of the PES. Inaccurate symmetry functions can treat
inequivalent configurations as equivalent, possibly facing the NN
with the task to fit two different input energies to ``nominally''
the same structure. Such contradictions typically give rise to bad
fits. Vice versa, truly equivalent structures should also yield the
same set of symmetry function values, since only then the symmetry
of the surface is exactly included in the PES representation (and in
the forces), and a numerical symmetry breaking affecting the MD
trajectories is impossible. Both requirements are met by the current
approach. It is also important to repeat that the mapping of the
coordinates is always done only in one direction: From the six
molecular coordinates to the 11 symmetry functions for the training
of the NN, as well as for the prediction of energies and forces of
new structures. It is not necessary to reconstruct the original set
of coordinates from the set of symmetry function values
$\{G^{\mu}\}$, which also allows us to use rather complicated symmetry
functions that are hard to invert.

An application of the here described Fourier method to other surface
unit-cell shapes and sizes, i.e. other single-crystal surfaces,
should in most cases be straightforward. For rectangular surface
unit-cells, e.g. at fcc(100) surfaces, the centers of three atomic
Fourier terms can e.g. be placed at the top, bridge and hollow site
to span the symmetry unique wedge of the surface. Also larger
surface supercells may readily be used, e.g. in the presence of
preadsorbed atoms or surface reconstructions, if the symmetry unique
wedge of the surface is chosen accordingly. Finally, in principle it
is also possible to describe the simultaneous interaction of two or
more diatomic molecules with the surface. In this case a set of
symmetry functions is constructed for each of the two molecules and
augmented by terms describing the relative position of the
molecules. However, in this case, as well as in the case of
poly-atomic molecules, the configuration space, i.e. the
dimensionality of the problem, is strongly increased making a
systematic mapping of the PES more costly. This is particularly
consequential, since it cannot be overstressed that NNs are not able
to extrapolate the energies of structures outside the configuration
spanned by the input data set. A systematic mapping of the entire
PES range of interest as done in the O$_2$/Al(111) model case is
thus a prerequisite for the presented approach.

\section{Summary}

We have presented a symmetry-adapted neural network representation
of potential-energy surfaces for molecule-surface interactions. The
method builds on symmetry functions, which fully take the symmetry
of the surface into account. Instead of the molecular coordinates,
the values of these symmetry functions are used as input for the
neural network. The symmetry functions are constructed in a
systematic way from atomic Fourier terms and the construction recipe
should be readily applicable to a wide range of surface unit-cell
sizes and shapes. The accuracy of the method has been demonstrated
by interpolating the six-dimensional potential-energy surface of an
oxygen molecule in the spin-triplet state interacting with the
Al(111) surface.

\section{Acknowledgements}

The authors wish to thank Matthias Scheffler for initiating the
neural network project and for stimulating discussions. Bernard
Delley is gratefully acknowledged for providing the DMol$^3$ code.


\begin{thebibliography}{99}
\bibitem{kalmandetails}
The employed parameters of the Kalman Filter are $\lambda$(0) =0.98000 and $\lambda_0$ =0.99770 following the description in Ref.~\onlinecite{lorenz06}.
\end{thebibliography}
\end{document}